\documentclass[10pt]{article}
\usepackage{times}
\usepackage{amsfonts}
\usepackage{graphicx}
\usepackage{float}

\begin{document}

\noindent {\Large \textbf{Brain reaction times: Linking individual
and collective behaviour through Physics modelling}}
%\\ \\
%Short Title: \\ \textbf{Brain Thermodynamics}}\\\\

 \vskip.5cm \noindent
{\bf J. C. Castro-Palacio}$^{1}$$^,$$^{a}$, {\bf P. Fern\'andez de C\'ordoba}$^{2}$$^,$$^{b}$,  {\bf  J. M. Isidro}$^{2}$$^,$$^{c}$,\\
and {\bf E. Navarro-Pardo}$^{3}$$^,$$^{*}$\\
${}^{1}$ Imperial College London, London SW7 2AZ, United Kingdom \\Current affiliation: Instituto Universitario de Matem\'atica Pura y Aplicada, Universitat Polit\`ecnica de Val\`encia, Valencia 46022, Spain\\
${}^{2}$ Instituto Universitario de Matem\'atica Pura y Aplicada, Universitat Polit\`ecnica de Val\`encia, Valencia 46022, Spain\\
${}^{3}$ Department of Developmental and Educational Psychology, Universitat de Val\`encia, Valencia 46010, Spain\\
${}^{a}${\tt juancas@upvnet.upv.es},
${}^{b}${\tt pfernandez@mat.upv.es}, \\ ${}^{c}${\tt joissan@mat.upv.es},\\ ${}^{*}${\tt Corresponding author: esperanza.navarro@uv.es}

\vskip.5cm \noindent \today \vskip.5cm \noindent

{\small \noindent

An individual's reaction time data to visual stimuli have usually
been represented in Experimental Psychology by means of an
ex-Gaussian function. In most previous works, researchers have
mainly aimed at finding a meaning for the parameters of the
ex-Gaussian function in relation to psychological phenomena. We will
focus on interpreting the reaction times of a group of individuals
rather than a single person's response, which is relevant for the
different contexts of social sciences. In doing so, the same model
as for the Ideal Gases (an inanimate system of non-interacting
particles) emerges from the experimental reaction time data. Both
systems are characterised by a collective parameter which is $k_BT$
in the case of the system of particles and what we have called
\textit{life span parameter} for the system of brains. Similarly, we
came across a Maxwell-Boltzmann-type distribution for the system of
brains which provides a natural and more complete characterisation
of the collective time response than has ever been provided before.
Thus, we are able to know about the behaviour of a single individual
in relation to the coetaneous group to which they belong and through
the application of a physical law. This leads to a new entropy-based
methodology for the classification of the individuals forming the
system which emerges from the physical law governing the system of
brains. To the best of our knowledge, this is the first work in the
literature reporting on the emergence of a physical theory (Ideal
Gas) from human response time experimental data.}

\newpage

\noindent The literature relates an important number of works on
human reaction times to visual stimuli
\cite{nature3,nature2,science1,pnas2,nature1,nature4}. This is a
very common scenario in daily life and also in a broad plethora of
situations in industry \cite{ind1}, behavioural economics and
finances \cite{bec1}, sports \cite{sports}, and health
\cite{health1}, to mention just a few examples. It is the collective
behaviour of a group of individuals that is really relevant in these
contexts.

\noindent In Experimental Psychology, the response time (RT) data of
an individual are usually represented in terms of the ex-Gaussian
function which provides three parameters, denoted as $\mu$,
$\sigma$, and $\tau$, coming from the convolution between an
exponential and a Gaussian function. Response time distributions are
positively skewed \cite{r14,GAMER2018} and so are not properly
described by standard central tendency estimators, such as the mean
and standard deviation \cite{r16}. In this respect, the ex-Gaussian
function has been proven to optimally fit the probability
distribution curves of the RT.

\noindent The ex-Gaussian parameters derived from representing RT
data are usually interpreted in connection with cognitive disorders
as certain correlations are claimed to exist
\cite{r17,r18,espe1,espe2,espe3,espe4,espe5}. For instance, the most
commonly diagnosed cognitive disorder in childhood affecting the RT
is Attention Deficit and Hyperactivity Disorder (ADHD). In this
respect, the exponential component $\tau$ has been found to
correlate with ADHD specifically in male children \cite{r23}. In
reference \cite{r22}, the authors studied ADHD and autism spectrum
disorder (ASD) in children aged 7-10 years to gain insights into the
attentional fluctuations, related to increased response time
variability.  The use of the ex-Gaussian function to fit reaction
time data has been a widespread procedure in the literature
\cite{r14,GAMER2018,r16,r17,r18,espe1,espe2,espe3,espe4,espe5,r23,r22,r21,MDPI2019}.

\noindent When we review previous works, it can be said that a
clear, direct and explicit regularity from the correlation between
the ex-Gaussian parameters and cognitive disorders has not yet been
reported, such that useful tools e.g. for mental disorder diagnosis
or any other applications could be implemented. This work builds on
the analysis and interpretations carried out to date although with a
global and systemic view of the group of individuals. Our aim is in
line with the everlasting search for general regularities in the
living world \cite{pnas1}.

\noindent In this article we seek to get new insights into the
reaction time data by means of the collective response of a group of
individuals. We go from the ex-Gaussian function, commonly used to
parameterise the response time data of an individual, to the
emergence of a Maxwell-Boltzmann (MB)-type distribution representing
the collective behaviour of a group of individuals, namely, the
probability distribution of the individual responses within a group.
We discuss the context for the appearance of the Ideal Gas Theory
\cite{TOLMAN} and therefore introduce a one-to-one correspondence
between a system of brains and a system of particles.

\vskip.5cm

\noindent {\large \textbf{Experiments}}\label{EXP}
\\\\
\noindent \textbf{Participants}
\\\\
For this research, a sample of 168 children (84 males and 84
females) with ages between 8 and 10 years was taken. The children
were uniformly randomly chosen and the mean age was 9.1 years, with
a standard deviation of 0.9 years. The children who participated in
the experiments attended a Primary School in Valencia (Spain). All
children were healthy, namely, presented no seizures, brain injury
or any other neurological damage. This information was obtained from
parents and school psychologists, who assessed all the children
biannually. We obtained all necessary consents and authorisations at
all necessary levels, specifically, the management of the school,
and the Regional Education Authority. This study was carried out in
accordance with the recommendations of the Secretariat of Education
of the Valencian Community. The protocol was approved by the
Government of Valencia (Generalitat Valenciana). The due written
consent of the children's parents or legal guardians was obtained in
accordance with the Declaration of Helsinki \cite{helsinki}.

\vskip.5cm

\noindent \textbf{Visual stimuli}
\\\\
Computer-based experiments were carried out using the Windows
program DMDX \cite{DMDX2003} widely used in the community of
experimental and cognitive psychologists
\cite{espe1,espe3,espe4,MDPI2019,DMDX2003,garai,ras}. By means of
this program, stimuli were presented to the participants and RTs
recorded. The tasks applied in this work correspond to the child
version of the Attentional Network Test (ANT Child) \cite{posner3}.
Each child was sitting in a separate and quiet place such that the
independence among children while performing the experiments was
guaranteed. Laptops bearing DMDX software were used. Each experiment
lasted for 6--7 minutes and stimuli were presented randomly to avoid
order presentation effects. Besides, in this type of experiment, the
individual response times take place in a very short period of time,
typically within a few hundred milliseconds (70-2500 ms). \noindent
The Attention Network Tasks seek to test three attentional networks:
alerting, orienting, and executive control \cite{FAN2002}. Alerting
network is assessed by changes in reaction time as a result of a
warning signal. Orienting is related to changes in the RT indicating
where the target will take place. Finally, the efficiency of the
executive control is tested by asking the children to answer by
pressing the keys in indicating left or right direction of an image
placed at the centre in between neutral, congruent or incongruent
flankers. These three networks are very related among themselves
\cite{posner1,posner2}. The visual stimuli were very simple and
uncorrelated to any cultural background or educational training.
Specifically, each stimulus consisted of 5 fish aligned horizontally
looking to the right or the left. The colour was black and the
background was white. The objective of the task was to identify, in
each trial, the direction of the central fish. There were three
cases depending on the orientation of the fish around the central
one. The neutral case was when there was only the central fish. The
congruent case was where the surrounding fish were placed in the
same direction as the central fish. The third case was where the
surrounding fish were placed in the opposite direction in respect to
the central one. If the central fish was facing right, the key
labelled ``M" should be pressed; if not, the key labelled ``Z"
should be pressed. A total of 144 stimuli were presented in a random
way and for a maximum of 2500 ms or until the child pressed a key.
Overall, our experiments involved $144\times168=24192$ reaction
times.

\newpage

\noindent {\large \textbf{Statistics of the reaction time data}}\\\\
\noindent Let $i$ be the index specifying the $i$--th individual,
where $i=1,\ldots,N=168$. Every individual is submitted to a test
that yields a set of reaction times whose distribution can be
represented by an ex-Gaussian function. This distribution is
characterised by three parameters which can be expressed as a
three-dimensional vector,
\begin{equation}
\vec a_i = (\mu_i, \sigma_i, \tau_i)
\label{uno}
\end{equation}

\noindent Let $f_i(t)$ be the ex-Gaussian probability distribution
of the response times of the $i$--th individual \cite{GAMER2018}:
\begin{equation}
f_i(t)=f(\vec a_i; t)=f(\mu_i, \sigma_i, \tau_i; t)
\label{dos}
\end{equation}
where $f_i(t)$ is,
\begin{equation}
f_i(t)=
\frac{1}{2\tau_i}\exp\left(\frac{1}{2\tau_i}(2\mu_i+\frac{\sigma_i^2}{\tau_i}-2t)\right)erfc\left(\frac{\mu_i+\frac{\sigma_i^2}{\tau_i}-t}{\sqrt2\sigma_i}\right).\label{dos1}
\end{equation}
In the previous equation, $erfc$ is the complementary error
function. We will characterise the distribution of RTs of the
$i$--th individual through its moments which is a more general way
than using $\mu_i$, $\sigma_i$, and $\tau_i$. We can consider the
moments centred either at the origin (raw moments), or at the
corresponding average (central moments). Thus
\begin{equation}
M_i\equiv \int_{-\infty}^{\infty} t f_i(t)dt
\label{tres}
\end{equation}
is the raw moment of order one, while
\begin{equation}
S_i^2\equiv\int_{-\infty}^{\infty}(t-M_i)^2f_i(t){\rm d}t
\label{cuatro}
\end{equation}
is the variance, or second central moment, of the random variable
$t$ with probability distribution $f_i(t)$ and centred at $M_i$. The
positive square root of the above, $S_i$, is the standard deviation.
On the other hand, the skewness $\lambda_i$ of the distribution is
defined as
\begin{equation}
\lambda_i\equiv\frac{1}{S_i^2}\int_{-\infty}^{\infty}(t-M_i)^3f_i(t){\rm
d}t \label{cinco}
\end{equation}
that is, the centred moment of order three, divided by the standard
deviation squared. We have divided by $S_i^2$ instead of $S_i^3$ as
we want to keep the same time dimension (milliseconds) for all three
moments.

\noindent Specifically for the ex-Gaussian function, the
aforementioned moments have the following form: mean of the
distribution, $M = \mu + \tau$ (first moment, Eq. \ref{tres}), the
variance $S^2 = \sigma^2 + \tau^2$ (second moment, Eq.
\ref{cuatro}), and the skewness $\lambda =
2\tau^3/\left(\sigma^2+\tau^2\right)$ (third moment, Eq.
\ref{cinco}) \cite{GAMER2018}. Then, it is clear that $\mu$,
$\sigma$, and $\tau$ are not the moments of the distribution but
parameters of the Gaussian and exponential functions that convolute
to give rise to the ex-Gaussian distribution.
 Altogether, an ex-Gaussian distribution
can be characterised through its three moments
\begin{equation}
\vec b_i=(M_i, S_i, \lambda_i).\quad  \label{seis}
\end{equation}
\noindent We have considered the first three moments only as we want
to keep the dimensionality of the vector in Eq. \ref{seis} equal to
three. The experimental input is the set of all $N$ vectors $\vec
b_i$ representing the $N$ individuals analysed.

\begin{figure}[H]
\centering
\includegraphics[height=6.61755in,width=3.087in]{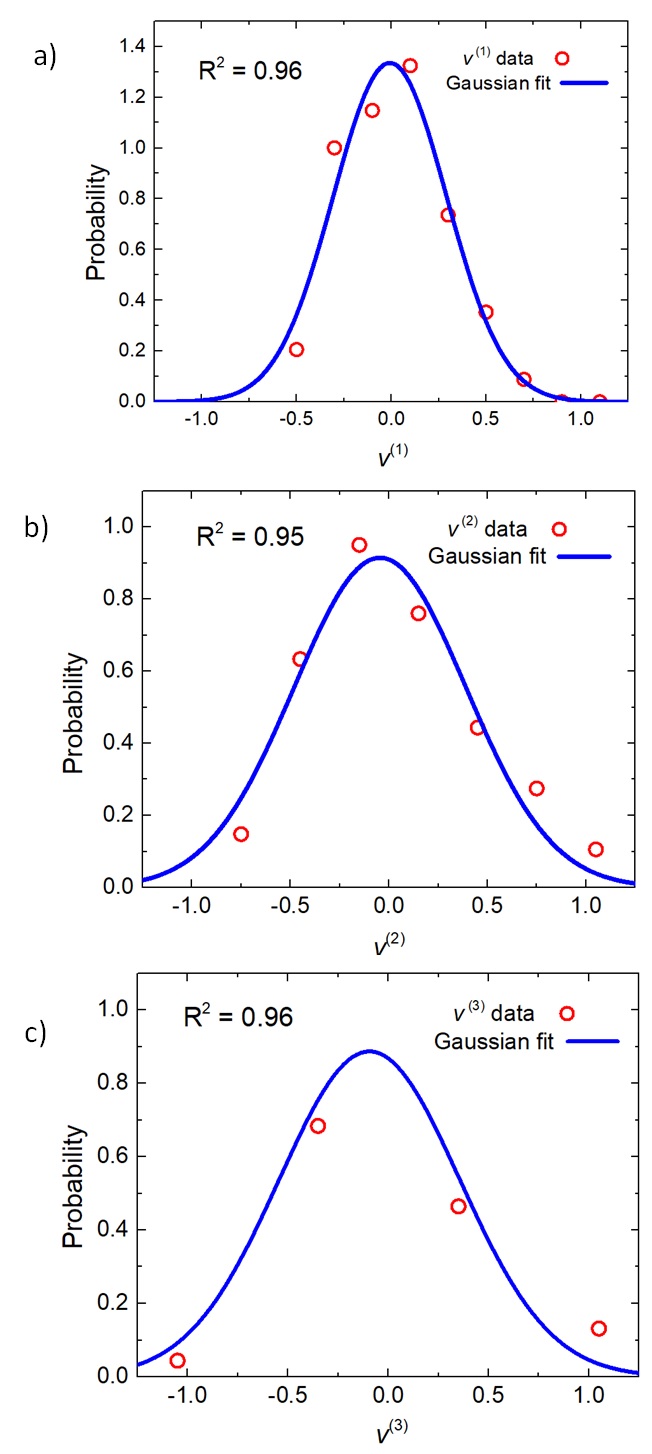}
\caption{Probability distributions of the dimensionless moments
calculated from $24192$ experimental reaction times over a sample of
168 children (red open circles). The mean is shown in panel a), the
standard deviation in panel b) and the skewness in panel c). The
$R^2$ coefficient of the fit (blue solid line) is also shown.}
\label{fig:fig1}
\end{figure}

\noindent Let us centre all vector components of $\vec b_i$ about
their means, that is,
\begin{equation}
(M_i-\bar M, S_i-\bar{S}, \lambda_i-\bar \lambda ), \quad
\label{siete}
\end{equation}
where
\begin{equation}
\bar M=\frac{1}{N}\sum_{i=1}^NM_i,\quad
\bar{S}=\frac{1}{N}\sum_{i=1}^NS_i,\quad \bar
\lambda=\frac{1}{N}\sum_{i=1}^N\lambda_i \label{ocho}
\end{equation}
Furthermore we will consider the vector $\vec v_i$ with dimensionless components
\begin{equation}
\vec v_i=(v_i^{(1)},v_i^{(2)},v_i^{(3)})\equiv\left(\frac{M_i-\bar
M}{\bar M},\frac{S_i-\bar{S}}{\bar{S}}, \frac{\lambda_i-\bar
\lambda}{\bar \lambda}\right). \quad \label{nueve}
\end{equation}
\noindent The three components of $\vec v_i$, that is, $v_i^{(1)}$,
$v_i^{(2)}$, and $v_i^{(3)}$, are evidenced to be distributed
normally, that is, they obey a Gaussian distribution over the
ensemble of $N$ individuals. The graphs in Figure \ref{fig:fig1}
(panels a), b), and c)) exhibit this property explicitly for each
component, respectively. At the 0.05 confidence level, the data of
the three moments were found to be significantly drawn from a
normally distributed data according to the Kolmogorov-Smirnov
normality test \cite{doob}. Therefore, the sample size used in this
work ($N = 168$) is enough to show the Gaussian behaviour of the
distributions of $v_i^{(1)}$, $v_i^{(2)}$, and $v_i^{(3)}$. In fact,
when we randomly choose a sample of 50 individuals out of 168, the
percentage relative errors among the respective variances of
$v_i^{(1)}$, $v_i^{(2)}$, and $v_i^{(3)}$ stay within 3.6 $\%$.

 \vskip.5cm

\begin{figure}[H]
\centering
\includegraphics[height=5.35808in,width=3.78248in]{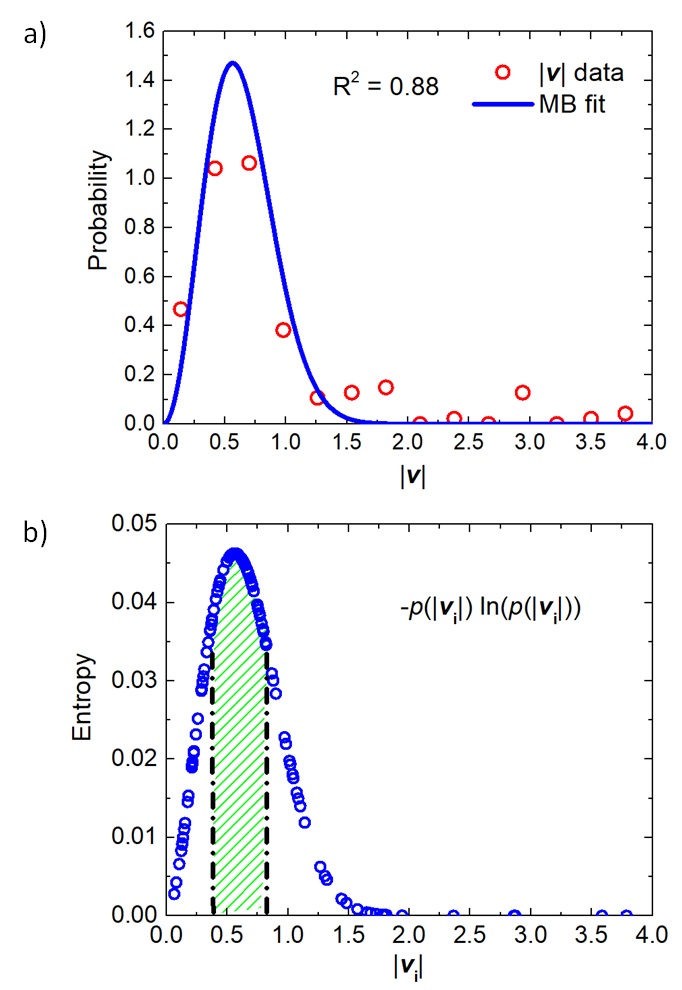}
\caption{Maxwell-Boltzmann-like distribution of the system of brains
in panel a) and the corresponding entropy in panel b). The data
shown in panel a) involved $24192$ experimental reaction times over
a sample of 168 children. The coefficient of determination of the
fit, $R^2$, is also shown. As an example, 50\% of the children have
been represented by a patterned region in panel b).}
\label{fig:fig2}
\end{figure}

\noindent {\large \textbf{Appearance of the MB distribution within the Ideal Gas theory}}\\\\
\noindent As a consequence of the foregoing, it is well known that
the modulus $\vert \vec v_i\vert$
\begin{equation}
\vert \vec v_i\vert\equiv
\sqrt{\left(v_i^{(1)}\right)^2+\left(v_i^{(2)}\right)^2+\left(v_i^{(3)}\right)^2}
\label{diez}
\end{equation}
obeys the Maxwell--Boltzmann probability density \cite{TOLMAN},
\begin{equation}
Q(x)=\sqrt{\frac{2}{\pi}}{B^{-3/2}}x^2\exp\left(-\frac{x^2}{2B}\right)
\label{once}
\end{equation}
where $x$ stands for the random variable $\vert \vec v_i\vert$, and
$B$ is the only parameter of the distribution $Q(x)$. Figure
\ref{fig:fig2} (panel a)) shows the fit of the Maxwell--Boltzmann
distribution (Eq. \ref{once}) to all the RT experimental data. The
fittings in Figures \ref{fig:fig1} and \ref{fig:fig2} were carried
out by the non-linear fitting algorithm of Levenberg--Marquardt
\cite{lev,mar}. The fitted value of $B$ is $ 0.159 \pm 0.010$
($R^2=0.88$). We will call the constant $B$ from now on the
\textit{life span parameter} ($LSP$) as this is the constant that
characterises the system of individuals at a given time of the life
span. The Gaussian behaviour of the distributions of $v_i^{(1)}$,
$v_i^{(2)}$, and $v_i^{(3)}$ leads to a Maxwell-Boltzmann
distribution of $\vert \vec v_i\vert$, even if in our case, the
variances of the three components are not all equal. As already
mentioned, we have obtained a coefficient of determination of $R^2 =
0.88$ using $24192$ experimental reaction times.

 \noindent
From the probability density function $Q(x)$ with random variable
$x$, a Boltzmann entropy can be defined \cite{TOLMAN}. Let's denote
it as ${\cal S}$,

\begin{equation}
{\cal S} = - \sum_{\rm {i=1}}^{\rm {N}} p_i \ln p_i \label{doce}
\end{equation}

\noindent where, in our case, N is the number of individuals in the
sample, and $p_i$ is the probability associated to the i-\textit{th}
individual,

\begin{equation}
p_i = Q(\vert \vec v_i\vert)/ \sum_{\rm {j=1}}^{\rm {N}}Q(\vert \vec
v_j\vert)
\end{equation}

 \newpage

\noindent This entropy satisfies an extremum property, namely,
imposing the constraint that $Q(x)$ has a given variance, the
Maxwell-Boltzmann distribution maximises the Boltzmann entropy
\cite{TOLMAN}. Figure \ref{fig:fig2} (panel b)) shows the entropy
curve for the experimental data of this work. Entropy is not only a
widely-used concept in Physics \cite{TOLMAN}, but also an analysis
tool in a variety of fields outside Physics
\cite{prado,collell,tsallis1,tsallis2}. Particularly in Psychology,
entropy has been applied, for instance, to analyse cognition
experiments \cite{prado,collell}.

\noindent In order to get access to the collective RT behaviour of a
group of individuals it is necessary to go up to a higher level of
description such as that provided by the Maxwell-Boltzmann
distribution which appears in this work from the experimental RT
data. Around the maxima of the MB distribution and of the entropy in
figure \ref{fig:fig2} panels a) and b), respectively, the most
likely individuals are located. This is very useful and relevant
information for the analysis in many contexts such as behavioural
economics or normative development in Psychology, where a central
aim is to get to know the behaviour of the group. At this point we
have made the transition from the description of the RTs of an
individual, expressed in terms of the $N$ vectors $\vec v_i$, to an
ensemble description in terms of the parameter $B$. On the other
hand, a single individual can be traced through to determine their
location at the entropy curve (figure \ref{fig:fig2} panel b)) in
order to check for their contribution to the total entropy of the
system. In this fashion, we can know about their behaviour in
relation to the coetaneous group they belong to. Actually, based on
the contribution of each individual to the entropy, a new
classification methodology could be established as the less and the
more entropic individuals inside a group can be distinguished. We
would like to point out that this entropy-based classification
methodology is not an arbitrarily-defined criterion but a
consequence of a behaviour existing in nature, which is one of the
most relevant findings of this work.

\noindent The system of $N$ individuals is basically characterised
by the parameter $B$ in Eq. \ref{once}, namely, the \textit{life
span parameter} ($LSP$). The $LSP$ plays a role analogous to that of
${\it k_BT}\/$ in the statistical theory of the Ideal Gas, where
$k_B$ is the Boltzmann constant and $T$ the temperature of the
particle system. As we have remarked in previous sections, our
sample consists of independent individuals. This leads to treating
them as an ideal gas, where particles do not interact among
themselves (only by elastic collisions). In this system, the number
of microstates accessible to each particle is proportional to the
phase space volume \cite{TOLMAN}. The logarithm thereof is
proportional to the entropy ${\cal S}$. The equation of state
follows immediately from here, that is, the First Law of
Thermodynamics states $\delta Q=dU+pdV$, and $\delta Q=Td{\cal S}$.
Differentiation of the Fundamental Thermodynamic Equation ($FTE$),
specifically ${\cal S}={\cal S}(U,V)$, yields $d{\cal S}=(\partial
{\cal S}/\partial U)dU+(\partial {\cal S}/\partial V)dV$. One infers
$T^{-1}=(\partial {\cal S}/\partial U)$ and $p=T\partial {\cal
S}/\partial V$. The latter is the equation of state, and yields
useful information if the $FTE$ is known. For the case of
independent particles, one has that ${\cal S}$ is proportional to
$\ln V$ up to the addition of some function $f(U)$. Hence $\partial
{\cal S}/\partial V$ is proportional to $1/V$, whence $pV$ is
proportional to $T$: the equation of state of an ideal gas.

\begin{figure}[H]
\centering
\includegraphics[height=3.34in,width=4.93in]{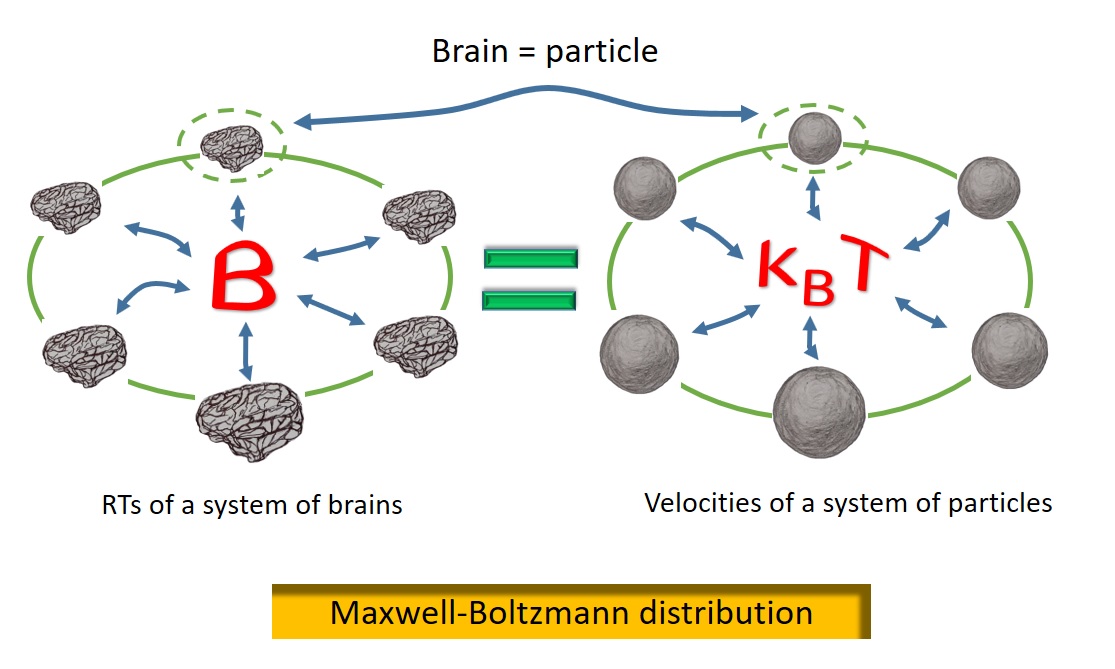}
\caption{Schematic representation of the correspondence between a
system of brains and a system of ideal gas particles. Both systems
are characterised by a Maxwell-Boltzmann distribution.}
\label{fig:fig3}
\end{figure}

\noindent Interestingly, there are clear evidences of a one-to-one
correspondence between the collective time response of a group of
children and a system of particles forming an ideal gas. Both
systems are characterised by the same model. In Figure
\ref{fig:fig3} the main elements of this correspondence are
depicted. The brain can be seen as a particle. The three moments of
the ex-Gaussian distribution for each child find their analogue in
the three components of the velocity vector of a particle. The
distribution of the modulus of a vector defined over the moments
leads to a chi distribution with three degrees of freedom which is
the Maxwell-Boltzmann distribution in the case of an ideal gas. The
particles form a system from their interaction with a thermostat
characterised by its temperature. Similarly, there is a
thermostat-like entity that makes the group of children be a system.
We can figure out this thermostat as a gene-driven entity of the
life span of the group of individuals determining their evolution
from birth to death in an analogous fashion as a physical thermostat
thermalises a particle system. We are moving to another level of
generality beyond the ex-Gaussians which characterise the RT of an
individual, in order to capture a regularity in the collective time
response of a group of children. These results represent a major
step in two directions. We can now know about the collective time
response of a group of individuals and also about the behaviour of a
single individual in relation to the coetaneous group they belong
to. These are the main contributions of this work.

\vskip.5cm
\noindent {\large \textbf{Conclusions}}\\\\

\noindent  In this work we discuss the appearance of the Ideal Gas
theory when describing the response time data of a group of
individuals. The ex-Gaussian function represents well the response
time data of an individual, but it fails to provide a complete and
comprehensive characterisation of the collective response time. The
collective response of the group can be only unveiled by going up to
a higher level of description such as that provided by a
Maxwell-Boltzmann distribution. To validate our model, we used
$24192$ experimental reaction times, obtaining a coefficient of
determination of $R^2 = 0.88$. One major step with respect to
previous works is that we can now know about the behaviour of a
single individual in relation to the contemporary group and in the
framework of a physical law. In practical terms, our proposal of an
entropy-based classification of individuals is not an \textit{ad
hoc} criterion but a consequence of a natural law.

\noindent The transition from the response time of an individual to
the collective response time of a group of individuals of the same
species comes from the fact that the individuals form a system by
means of connection with the thermostat-like entity which can be
considered gene-driven and thus being an entity related to the human
species. Then, a Maxwell-Boltzmann-type distribution (a chi
distribution) appears naturally from the experimental RT data. On
the grounds of the emergence of a MB distribution is the
experimental evidence that the moments of the ex-Gaussian
distributions follow Gaussian distributions among the group of
children. It should be pointed out that the Gaussian distribution of
the moments among the individuals is something driven by nature.

\noindent The literature includes a few works about Physics and
Psychology \cite{pp1900,pp1953,pp1995}, which are monographic essays
where the authors hold conceptual discussions about the
possibilities of using Physics to explain psychological phenomena.
Some of them develop models, for instance, that use quantum theory
to approach brain-related phenomena
\cite{dzh,acacio,bruza,graben,haven,khren}. Another group of works
use Physics concepts to understand psychological data, such as those
applied to understand cognition experiments \cite{prado,collell}.
What is really noteworthy in the present work is the natural
appearance of a physical theory in the context of human response
time experimental data. In fact, our results show that the
underlying fundamental model for the independent particles of an
ideal gas is the same as for the response times of a group of
individuals.

\vskip.5cm
\noindent {\large \textbf{Outlook}}\\\\

\noindent In upcoming works, we expect to gain further insight into
the existing correspondence between the response time of a group of
individuals and the Ideal Gas of independent particles.

\noindent In this work, a way of representing the collective
behaviour of a group of individuals is reported. This result has a
number of implications in different areas of knowledge, such as
education, mental disease diagnosis, decision making in behavioural
economics, and finances. We want to address questions such as the
ageing process troughout the life span, what is considered a
normative behaviour in education or health, the appearance of the
life span parameter ($LSP$) and its connection to the features of
the species, the philosophical implications of the existence of an
ideal gas law behaviour of the fast response times in humans, among
others.

\noindent We have proven that the reaction times of a group of
individuals of the same age obeys a Maxwell-Boltzmann distribution,
as it corresponds to an Ideal Gas of independent particles at a
given temperature. To count on a model able to describe the
collective behaviour at a given time, strongly suggests to work on a
model describing the evolution of the group across the life span.
Given that the group at a given time is characterised by what we
have called above as the \textit{life span parameter}, which is
analogous to $k_BT$ in an ideal gas, the life span can then be
modelled as a continued evolution (thermalisation) with evolving
values of $LSP$ and entropy until the system reaches a final
equilibrium with a thermostat-like entity. We are also interested in
analysing the collective behaviour of other variables than reaction
time not only in humans but also in other living beings.

\vskip.5cm
\noindent {\large \textbf{Acknowledgements}}\\\\
\noindent Authors would like to thank Prof. Miguel \'Angel Garc\'ia
March, Prof. Daniele Tommasini, Prof. Alberto Conejero, Prof.
Luisberis Vel\'azquez, and Dr. Ismael Orqu\'in for the fruitful
discussions about the manuscript. The psychologists from the
Interdisciplinary Modelling Group InterTech (www.intertech.upv.es)
who participated in the collection of the experimental data are
acknowledged in this work. Imperial College London is also
acknowledged as it was the institution where J.C.C.P. carried out
the majority of his contribution to this article. This research was
partially supported by grant number RTI2018-102256-B-100 (Spain).

\vskip.5cm

\noindent {\large \textbf{Author Contributions}}\\\\
\noindent E.N.P. developed the psychological part of the paper and
carried out the experiments; E.N.P. and P.F.-d.-C. collected the
data; all authors analysed the data and developed the model; J.M.I.
and P.F.-d.-C. wrote most of the methodological aspects of the
paper; J.C.C.P. did the statistics, produced the figures, and wrote
the manuscript except the methodological aspects; J.M.I., E.N.P.,
and P.F.-d.-C. made written comments on different parts of the
manuscript; all authors reviewed and approved the final manuscript.

\vskip.5cm
\noindent {\large \textbf{Additional Information}}\\\\
\noindent \textbf{Competing Interests}: The authors declare no
competing interests.

\end{document}